\documentclass[reprint,amsmath,amssymb,aps,prb,superscriptaddress, showpacs, noeprint]{revtex4-2}
\usepackage{amsmath,amssymb,ascmac,fancybox}
\usepackage{color}
\usepackage{graphicx}
\usepackage{url}
\usepackage{siunitx}
\usepackage{bm}
\usepackage{physics}
\usepackage{mathtools}
\usepackage{hyperref}
\usepackage[T1]{fontenc}
\usepackage{ulem}

\usepackage[whole]{bxcjkjatype}

\newcommand{\kB}{k_{\mathrm{B}}}
\newcommand{\ani}[1]{\hat{#1}}
\newcommand{\cre}[1]{\hat{#1}^\dagger}
\newcommand{\ene}{\varepsilon}
\newcommand{\bc}{\xi} 
\newcommand{\bcurv}{\mathcal{F}}
\renewcommand{\vec}{\vb*}
\newcommand{\lowband}{\mathcal{LB}}
\newcommand{\highband}{\mathcal{HB}}
\newcommand{\Heff}{H_{\mathrm{eff}}}
\newcommand{\Jeff}{J_{\mathrm{eff}}}

\newcommand{\omegaoutput}{\omega_{J}}

\newcommand{\proj}{\Pi}

\newcommand{\Gfunc}{\mathcal{G}}
\newcommand{\GR}{G^{\mathrm{R}}}
\newcommand{\GA}{G^{\mathrm{A}}}
\newcommand{\selfene}{\Sigma}
\newcommand{\selfeneR}{\selfene^{\mathrm{R}}}
\newcommand{\selfeneA}{\selfene^{\mathrm{A}}}

\newcommand{\Emat}{\mathcal{E}}

\newcommand{\sgn}{\mathrm{sgn}}

\begin{document}

\title{
Effects of relaxation on photovoltaic effect and possibility of photocurrent within transparent region 
}

\author{Yugo Onishi}
\affiliation{Department of Applied Physics, The University of Tokyo, 
Tokyo, 113-8656, Japan}

\author{Hikaru Watanabe}
\affiliation{RIKEN Center for Emergent Matter Science (CEMS), Wako, 
Saitama, 351-0198, Japan}

\author{Takahiro Morimoto}
\affiliation{Department of Applied Physics, The University of Tokyo, 
Tokyo, 113-8656, Japan}
\affiliation{JST, PRESTO, Kawaguchi, Saitama, 332-0012, Japan}

\author{Naoto Nagaosa}
\affiliation{Department of Applied Physics, The University of Tokyo, 
Tokyo, 113-8656, Japan}
\affiliation{RIKEN Center for Emergent Matter Science (CEMS), Wako, 
Saitama, 351-0198, Japan}

\date{\today}
\begin{abstract}
We theoretically study photocurrents in metals that break both inversion $\mathcal{P}$ and $\mathcal{T}$ symmetries within the transparent region. 
We find that the system under the ac electric fields is well described with an effective Hamiltonian and the photocurrent is of the order of $\order{\omegaoutput/\gamma}$ if the frequency of the induced current $\omegaoutput$ and the scattering rate $\gamma$ satisfy $\omegaoutput/\gamma \ll 1$, and vanishes in the limit of $\omegaoutput/\gamma\to 0$.
On the other hand, the effective Hamiltonian description indicates that  nonvanishing photocurrent can appear even in the transparent region if the system is thin enough compared to the mean free path in the direction of the induced current (where $\gamma$ can be effectively regarded as 0).
Candidate materials for the such photovoltaic effect within the transparent region include multiferroics breaking both $\mathcal{P}$ and $\mathcal{T}$ symmetries.
\end{abstract}

\maketitle

\section{Introduction}
Photovoltaic effect attracts recent intensive interests from 
the viewpoints of both fundamental physics and applications~\cite{VonBaltz1981, Sturman1992, Sipe2000, Young2012, Young2012a, Sotome2019, Sotome2019a, Sotome2021, Cook2017, Rangel2017, Zhang2018, Morimoto2016a, Nagaosa2017, Matsyshyn2021}.
It is now realized that the bulk photovoltaic effects are closely related to the geometric aspects 
of electronic states in solids (such as Berry phase~\cite{Berry1984}) 
that arise from
the breaking of the inversion symmetry $\cal{P}$ in noncentrosymmetric materials.
Shift current and injection current are the two major mechanisms
for the photovoltaic effect in the second order in the electric field of light. 
Time-reversal symmetry $\cal{T}$
is another important symmetry;
the injection current requires   
the broken $\cal{T}$ either by the circularly polarized light or the 
magnetic field/order of the material, 
whereas the shift current does not require the breaking of $\cal{T}$.
When both $\cal{P}$ and $\cal{T}$ are broken, 
the energy dispersion of the electrons becomes asymmetric
between $k$ and $-k$, and many nonreciprocal phenomena 
are expected including the magnetochiral anisotropy in dc transport~\cite{Tokura2018}. 
Also insulators with broken $\cal{P}$ and $\cal{T}$ are called 
multiferroics, which shows the strong coupling between the 
electric and magnetic degrees of freedom such as the magnetoelectric
effect~\cite{Mostovoy2006, Fiebig2005, Tokura2014}. 
It has been predicted that the excitation of the  electromagnon by light in multiferroics can induce the dc current;  the low energy light below the band gap can show the photovoltaic effect without the electronic particle-hole excitation ~\cite{Morimoto2019,Morimoto-Kitamura-Okumura21}.
Furthermore, it has been shown that the soft phonon excitation
in ferroelectric BaTiO$_3$ without $\cal{T}$-breaking produces the dc shift current
~\cite{Okamura2022}. 
A common feature of all these photovoltaic effect is that 
the absorption of light occurs in some form even if
the electronic excitation is virtual. Then, the crucial question here
is if the photovoltaic effect is possible even without the absorption of
light or not.

According to the standard perturbation theory~\cite{Parker2019, DeJuan2020, Watanabe2021, Sodemann2015, Gao2021}, it seems possible that, if both $\cal{T}$ and $\cal{P}$ symmetries are broken, finite photocurrent is induced even in the transparent region where the incident photon cannot create real excitations of electrons. Histrically, however, Belinicher {\it et al.} pointed out that there should be no dc photocurrent for the transparent region in the steady state if we properly take into account the effect of the relaxation~\cite{Belinicher1986}. The effect of relaxation on the photocurrent was discussed based on an effective Hamiltonian decades ago and they concluded that there should be no photocurrent in the steady state~\cite{Belinicher1986, Ivchenko1988}. However, since their discussion is based on the semiclassical theory, it is not clear how the photocurrent in the presence of relaxation is affected by the interband matrix elements of the current operator, which can be crucial in the photocurrent such as the shift current.

In Ref.~\cite{DeJuan2020}, it was also pointed out that the relaxation affects the photocurrent when the relaxation rate is larger than the frequency of the output current, although they studied only the region where the relaxation rate is much smaller than the frequency of the output current.

On the other hand, recent studies suggest that there is a possibility of finite photocurrent even in the transparent region \cite{Shi2022arXiv, Golub2022arXiv}. In Ref.~\cite{Shi2022arXiv}, a system coupled with particle reservoirs is discussed and it is concluded that the photocurrent within the transparent region exists in such systems and their relation to the thermodynamics is also discussed. In Ref.~\cite{Golub2022arXiv}, Golub {\it et al.} discussed a Raman-like process utilizing impurity scattering. In this case, the frequency of the incident photon and that of the scattered photon are different, resulting in finite energy absorption to drive the finite current.  
Since these studies discussed different situations or processes from the ones in Ref.~\cite{Belinicher1986, Ivchenko1988}, it is important to extend the previous discussions in Ref.~\cite{Belinicher1986,Ivchenko1988} and investigate the behavior of the photocurrent in the transparent region by incorporating relaxation effects and the interband matrix elements of the current operator.

In this paper, we perform analyses with a fully quantum mechanical theory on the photocurrent within the transparent region in the presence of relaxation effects.
As discussed previously in Ref.~\cite{DeJuan2020}, there are two important frequencies (or equivalently time scales) that determine the photocurrent responses: the relaxation rate $\gamma$ and the difference of injected and emitted lights frequencies which is equal to the frequency of current $\omegaoutput$. We usually consider the limit $\gamma\to 0$ and $\omegaoutput\to 0$, and their order is crucial to understand the photocurrent behavior. 
According to our theory, the photocurrent will vanish in the limit of $\omegaoutput/\gamma\to 0$, which is consistent with the previous discussions~\cite{Belinicher1986, Ivchenko1988}, while if one can realize the condition $\omegaoutput\gg\gamma$ effectively, nonvanishing photocurrent can appear. In this work, we often call the former case as ``slow limit'', while the latter case is denoted as ``fast limit''. Our theory also enables us to calculate the crossover at $\gamma\sim\omegaoutput$.  We also briefly discuss the interpretation of the photocurrent in the fast limit as a Raman process. While phonon excitation and structural phase transition induced through the virtual excitation of electrons were discussed previously~\cite{Zhou2020, Zhou2021}, we show that a similar Raman process without phonons can leave an electronic excitation with a finite current in the fast limit. A similar Raman-like mechanism of photocurrent is proposed by Golub {\it et al.} recently~\cite{Golub2022arXiv}, but our interpretation does not involve impurity scattering in contrast to Ref.~\cite{Golub2022arXiv}. In terms of applications of the bulk photovoltaic effect to solar cells, if the frequency of the output current is finite but sufficiently small, it is possible to rectify the current to extract power. Therefore, our results show a new possibility of energy harvesting of light within the transparent region.

In general, it is a complicated task to compute nonlinear responses taking into account the effect of scattering. Although one of previous studies~\cite{Rostami2021} investigated the effect of scattering on the nonlinear optical responses including vertex correction, we employ another approach to this problem. 
Specifically, we consider an effective Hamiltonian for the system in the presence of an ac electric field with frequencies smaller than the band gap. The perturbation in the effective Hamiltonian due to the applied field consists of two terms: one is a linear order term with respect to the external field and contains only an intraband contribution. This term does not contribute to the second order response in the clean limit, and thus we neglect it. The other term is second order in the external field and contributes to the second order response. 
In particular, this effective Hamiltonian description allows us to use the standard linear response theory to study the second order responses to the external field and facilitates to include the effects of scattering in the calculation. 
A similar approach of an effective Hamiltonian but with the Boltzmann equation was used in Ref.~\cite{Belinicher1986}. However, it is not clear in their approach how the interband matrix elements of the current operator can affect their results. We show that the effective current operator is given by the $k$-derivative of the effective Hamiltonian and extend the discussion in Ref.~\cite{Belinicher1986}.

This paper is organized as follows. In Sec.~\ref{sec:review}, we briefly review the known results from the standard perturbation theory in the clean limit~\cite{Parker2019, DeJuan2020, Watanabe2021, Sodemann2015, Gao2021}. In Sec.~\ref{sec:photocurrent_response}, we show the results for the photocurrent response within the transparent region and explain the relation with the previous studies. Sec.~\ref{sec:discussion} is devoted to discussion and conclusion.

\begin{figure*}
    \centering
    \includegraphics[width=1.8\columnwidth]{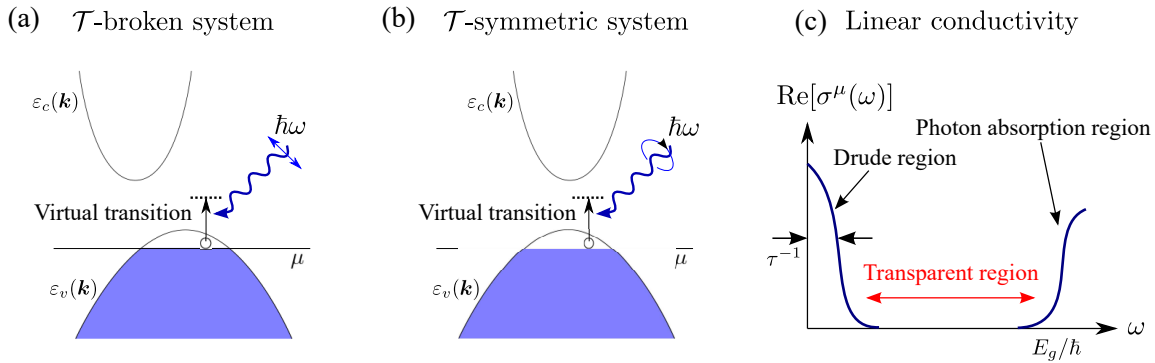}
    \caption{Schematic illustration of the photovoltaic effect within the transparent region. In this paper, we consider cases where (a) $\mathcal{T}$-broken systems under irradiation of light or (b) $\cal{T}$-symmetric systems under illumination of a circularly polarized light. According to the standard perturbation theory in the clean limit, even in these cases finite photocurrent can be induced, but the effects of relaxation can suppress the photocurrent. In these cases we clarify the condition for the photocurrent to be finite.
    (c) Illustration of $\omega$ dependence of linear conductivity $\sigma(\omega)$. At low frequencies $\omega < \tau^{-1}$ with the relaxation time $\tau$, the Drude response should be observed. At high frequencies comparable to or higher than $E_g/\hbar$ with the band gap $E_g$, photon absorption occurs and again $\Re[\sigma^\mu(\omega)]$ becomes finite.}
    \label{fig:schematics}
\end{figure*}

\section{Brief review of the standard perturbation theory} \label{sec:review}
In this section, we briefly review the standard perturbation theory in the clean limit described in the literature~\cite{Parker2019, DeJuan2020, Watanabe2021, Sodemann2015, Gao2021}.
Let us consider general noninteracting electronic systems with a periodic potential described by a Hamiltonian of the following form,
\begin{align}
	\hat{H}_0 &= \int\frac{\dd^d k}{(2\pi)^d}\cre{\psi}_{\vec{k}} H_0(\vec{k}) \ani{\psi}_{\vec{k}}, \label{eq:Ham_wo_field}
\end{align}
where $\ani{\psi}_{\vec{k}}, \cre{\psi}_{\vec{k}}$ are the annihilation and creation operators of an electron with wavevector $\vec{k}$, $H_0(\vec{k})$ is the Bloch Hamiltonian, and $d$ is the spatial dimension. The matrix $H_0(\vec{k})$ is hermitian and can be diagonalized with a unitary matrix $U_{\vec{k}}$ as 
\begin{align}
    \Emat_{\vec{k}} &= U_{\vec{k}}^\dagger H_0(\vec{k}) U_{\vec{k}},
\end{align}
where $(\Emat_{\vec{k}})_{ab} = \delta_{ab}\ene_{\vec{k}a}$ is a diagonal matrix, and $\ene_{\vec{k}a}$ is the energy dispersion for the band $a$.

The second order current responses to ac electric fields are described by the third-rank tensor $\sigma^{\mu\alpha\beta}$ defined as 
\begin{align}
    J^\mu(\omega_1 + \omega_2) &= \sigma^{\mu\alpha\beta}(\omega_1 + \omega_2; \omega_1, \omega_2)E^{\alpha}(\omega_1) E^{\beta}(\omega_2).
\end{align}
with the Fourier component of the electric current and the electric field, $\vec{J}(\omega)$ and $\vec{E}(\omega)$.  
In particular, the second order dc current response to ac electric fields is expressed as $\sigma^{\mu\alpha\beta}(0;\omega, -\omega)$. 
We can calculate the tensor $\sigma^{\mu\alpha\beta}(0;\omega, -\omega)$ with the standard perturbation theory~\cite{Parker2019, DeJuan2020, Watanabe2021, Sodemann2015, Gao2021}. 
In most literature, the current response for $\omega$ which is equal to or larger than the band gap was studied; for example, the shift current and the injection current with photon energy above the band gap are well understood with the perturbation theory. 
However, here we focus on responses to external fields in the transparent region. Namely, we assume the frequency of the external fields $\omega$ satisfies
\begin{align}
    \frac{1}{\tau}\ll \omega < E_g \label{eq:omega_condition}
\end{align}
where $\tau$ is the scattering time, and $E_g$ is the band gap of the system. We set $\hbar=1$ here and hereafter. For $\omega$ satisfying Eq.~\eqref{eq:omega_condition}, the external field cannot induce real electronic excitation, and also we can neglect the effects of Drude-like absorption.
In this case, there appears no delta-function term expressing the energy conservation and the absorption of light in the expression of the tensor, but there are still nonzero terms in the clean limit in general, which can be written as 
\begin{align}
    &\sigma^{\mu\alpha\beta}(0;\omega, -\omega)  \qq{(clean limit)} \nonumber \\
    &= \frac{e^3}{2\hbar^3\omega^2} \int\frac{\dd^d k}{(2\pi)^d} \sum_{a}(-\partial_\mu f_a) \nonumber \\
    & \quad \times \qty[h^{\alpha\beta}_{aa} + \sum_{b(\neq a)}\qty(\frac{h^{\alpha}_{ab}h^{\beta}_{ba}}{\ene_{ab}-\omega} + \frac{h^{\beta}_{ab}h^{\alpha}_{ba}}{\ene_{ab}+\omega})], \label{eq:perturbation_results}
\end{align}
with $\ene_{ab} = \ene_{\vec{k}a}-\ene_{\vec{k}b}$ and the Fermi distribution function $f_a = (e^{(\ene_{\vec{k}a}-\mu)/\kB T} + 1)^{-1}$~\cite{Parker2019, DeJuan2020, Watanabe2021, Sodemann2015, Gao2021}. Here, $\kB$ is the Boltzmann constant, $e(<0)$ is the charge of an electron, $T$ is the temperature and $\mu$ is the chemical potential. $\partial_\alpha$ is the derivative with respect to $k_{\alpha}$, and $h^\alpha, h^{\alpha\beta}$ are defined as 
\begin{align}
    h^{\alpha} &= U^{\dagger}_{\vec{k}} (\partial_{\alpha}H_0(\vec{k})) U_{\vec{k}}=\partial_{\alpha}\Emat_{\vec{k}} - i\comm{\bc^{\alpha}}{\Emat_{\vec{k}}}, \\
    h^{\alpha\beta} &= U^{\dagger}_{\vec{k}} (\partial_{\alpha}\partial_{\beta} H_0(\vec{k})) U_{\vec{k}} = \partial_{\alpha}h^{\beta} - i\comm{\bc^{\alpha}}{h^{\beta}}.
\end{align}
where $\bc^{\alpha}$ is the interband Berry connection defined by 
\begin{align}
    \bc^{\alpha} &= U^{\dagger}_{\vec{k}}\partial_{\alpha}U_{\vec{k}}.
\end{align}
$h^{\alpha}$ corresponds to the velocity operator in the band basis. We note that $\ene_{ab}, f_{a}, h^{\alpha}$ and $h^{\alpha\beta}$ are all $\vec{k}$-dependent although we suppress the indices $\vec{k}$ from them for shorthand notation. 

Here we emphasize that the result in Eq.~\eqref{eq:perturbation_results} is valid only in the clean limit, i.e., without any scattering or relaxation. If one properly take into account the effects of relaxation, the photocurrent vanishes with an additional contribution from the relaxation~\cite{Belinicher1986, Ivchenko1988}. In the following, we will clarify when the photocurrent in the transparent region is possible. 

\section{Photocurrent response} \label{sec:photocurrent_response}
In this section, we study the photocurrent response when the chemical potential crosses only one band and that band is well separated from the other bands energetically. Let us consider the case where the system described by Eq.~\eqref{eq:Ham_wo_field} is subjected to the ac electric field $\vec{E}(t)$. The Bloch Hamiltonian in the band basis is no longer diagonal due to the external field, and is given by 
\begin{align}
	H(t) &= \Emat - eA^\alpha(t) h^\alpha + \frac{e^2}{2}A^{\alpha}(t) A^{\beta}(t) h^{\alpha\beta} + \order{A^3}, 
\end{align}
where we omit the dependence of wavevector $\vec{k}$ in $H(t)$ and $\Emat_{\vec{k}}$. Here, $\vec{A}(t)$ is the vector potential and related to the electric field $\vec{E}(t)$ as $\vec{E}(t)=-\dv{\vec{A}}{t}$. We also assume that $\vec{A}(t)$ is of the form 
\begin{align}
    \vec{A}(t) &= \vec{a}(t)e^{-i\omega t} + (\vec{a}(t))^{*} e^{i\omega t},
\end{align}
where $\vec{a}(t)$ is slowly varying compared to frequency $\omega$.

In this case, the effective Hamiltonian $\Heff$ and the effective current operator  $\Jeff$ for the band near the chemical potential are given by 
\begin{align}
	\Heff(t) &= \ene_a - eA^{\alpha}(t)h^{\alpha}_{aa} + F^{\alpha\beta}b^{\alpha\beta}(t), \label{eq:Heff} \\
	F^{\alpha\beta} &= e^2 h^{\alpha\beta}_{aa} + e^2\sum_{b\neq a} \qty[\frac{h^{\alpha}_{ab}h^{\beta}_{ba}}{\ene_{ab}-\omega} + \frac{h^{\beta}_{ab}h^{\alpha}_{ba}}{\ene_{ab}+\omega}],\\
	\Jeff^{\mu}(t) &= \partial_\mu \Heff(t) + \pdv{P_{\mathrm{tr}}^{\mu}}{t}, \label{eq:Jeff_maintext} \\
	A^\alpha(t) &= a^{\alpha}(t) e^{-i\omega t} + (a^{\alpha}(t))^{*} e^{i\omega t}, \\
	b^{\alpha\beta}(t) &= a^{\alpha}(t)(a^{\beta}(t))^{*},
\end{align}
where $a$ is the band index for the band we are focusing on. In the rest of this paper, we often omit the band index $a$. The Einstein convention for $\alpha,\beta$ is always used in this paper.

We can confirm that, in the clean limit, the second term in the effective Hamiltonian (Eq.~\eqref{eq:Heff}) does not contribute to the second order dc current response and hence we ignore it hereafter. If the second term is neglected, Eq.~\eqref{eq:Heff} coincides with the effective Hamiltonian in Ref.~\cite{Belinicher1986}. 

The last term in the effective current operator, Eq.~\eqref{eq:Jeff_maintext}, is a total derivative with respect to time and hence it gives a transient current when the external field is applied. This term is closely related to the Berry connection and may be interpreted as the polarization current~\cite{Resta1994, Vanderbilt1993}. Please see Appendix \ref{ap:effective_Hamiltonian} for more details. In the rest of the present paper, we will ignore this transient contribution unless otherwise noted.
The rest of the effective current operator is given by $k$-derivative of the $\Heff$, and hence the application of $b^{\alpha\beta}(t)$ also induces the modification of the current operator as:
\begin{align}
    \Jeff^{\mu}(t) &= \partial_\mu \ene_a + b^{\alpha\beta}(t)\partial_\mu F^{\alpha\beta}.
\end{align}
We emphasize that, our derivation of $\Jeff^{\mu}(t)$ takes into account the interband matrix element in the original current operator. These matrix elements appear in $\Jeff^\mu(t)$ only through the $k$-derivative of $F^{\alpha\beta}$. For the derivation of the effective Hamiltonian and the detailed expression for $P_{\mathrm{tr}}$, see Appendix \ref{ap:effective_Hamiltonian}.

Several properties of $F^{\alpha\beta}$ should be noted. First, in terms of the time reversal symmetry $\mathcal{T}$, $F^{\alpha\beta}$ satisfies the following relationship if the system preserves $\mathcal{T}$:
\begin{align}
    F^{\alpha\beta}(\vec{k}) &= F^{\beta\alpha}(-\vec{k}),
\end{align}
where we explicitly show the $\vec{k}$-dependence of $F^{\alpha\beta}$. We also note that $F^{\alpha\beta}$ can be rewritten as 
\begin{align}
    F^{\alpha\beta} &= e^2 (\partial_{\alpha}\partial_{\beta}\ene_a) - ie^2\omega\epsilon_{\alpha\beta\gamma}\bcurv^\gamma \nonumber \\
    & \quad + e^2\omega^2\sum_{b\neq a} \qty(\frac{\bc^\alpha_{ab}\bc^\beta_{ba}}{\ene_{ab}-\omega} + \frac{\bc^\beta_{ab}\bc^\alpha_{ba}}{\ene_{ab}+\omega}). \label{eq:results2} 
\end{align}
Here $\bc^{\alpha}$ is the interband Berry connection and $\bcurv^{\alpha}=i\epsilon_{\alpha\beta\gamma}\sum_{b(\neq a)}\bc^{\beta}_{ab}\bc^{\gamma}_{ba}$ is the Berry curvature of the band $a$, while $b$ is a label for the other bands separated by the band gap. $\epsilon_{\alpha\beta\gamma}$ is the Levi-Civita symbol. This form also appears in the expression for dynamical Stark shift~\cite{Pershoguba2022arXiv} as discussed later in Sec.~\ref{sec:discussion}.

Now let us consider the linear response of the electric current to $b^{\alpha\beta}(t)$ to calculate the photocurrent response $\sigma^{\mu\alpha\beta}(0;\omega, -\omega)$. Please note that $b^{\alpha\beta}(t)$ is roughly related to the slow component of $E^{\alpha}(t)E^{\beta}(t)$ as
\begin{align}
    b^{\alpha\beta}(t) + b^{\beta\alpha}(t) \sim \frac{1}{\omega^2} \textrm{[slow component of }E^{\alpha}(t)E^{\beta}(t)]. \label{eq:b_EE_relation}
\end{align}
To be more precise, the frequency of the $b^{\alpha\beta}(t)$  corresponds to the frequency difference of the incident and scattered photons, which is equal to that of the output current, and hence we denote the frequency of $b^{\alpha\beta}(t)$ as $\omegaoutput$. 
Then we can denote the current response to $b^{\alpha\beta}$ as
\begin{align}
    J^{\mu}(\omegaoutput) &= \Phi^{\mu\alpha\beta}(\omegaoutput) b^{\alpha\beta}(\omegaoutput),
\end{align}
where $J^{\mu}(\omegaoutput)$ is the Fourier transformation of the current in $\mu$ direction and similar for $b^{\alpha\beta}(\omegaoutput)$. In the limit of $\omegaoutput\to 0$, $\Phi^{\mu\alpha\beta}$ corresponds to the dc photocurrent response $\sigma^{\mu\alpha\beta}(0;\omega, -\omega)$.
We can compute $\Phi^{\mu\alpha\beta}(\omegaoutput)$ with the standard linear response theory and the result in the imaginary time domain is written as
\begin{align}
	& \Phi^{\mu\alpha\beta}(i\omegaoutput) = \chi^{\mu\alpha\beta}(i\omegaoutput) + \int[\dd{k}] \expval{e\partial_\mu F^{\alpha\beta}}_0, \label{eq:response_func}\\
	& \chi^{\mu\alpha\beta}(i\omegaoutput) \nonumber 
	\\ &= \sum_n \int[\dd{k}] \Gamma^{\mu}(i\ene_n + i\omegaoutput, i\ene_n)\Gfunc(i\ene_n + i\omegaoutput)F^{\alpha\beta}\Gfunc(i\ene_n), \label{eq:chi_gfunc_expression}
\end{align}
where $i\omegaoutput, i\ene_n$ are Matsubara frequency, $\Gamma^{\mu}$ is the vertex for the current operator, and $\Gfunc(i\ene)$ is given by
\begin{align}
	\Gfunc(i\ene) &= \frac{1}{i\ene - \xi_{\vec{k}} - \selfene(i\ene, \vec{k})}.
\end{align}
$\xi_{k}$ is the energy dispersion of electron with respect to the chemical potential, and $\selfene$ is 
the self energy. $[\dd{k}]$ is shorthand for $\dd^d{k}/(2\pi)^d$ with the spatial dimension $d$.
$\expval{\dots}_0$ denotes the expectation value in the  equilibrium, and the last term in Eq.~\eqref{eq:response_func} corresponds to the diamagnetic current like contribution arising from the modification of the current operator due to the external field $b^{\alpha\beta}$, as we already mentioned.
In the above expressions, we omit the $k$-dependence of $\Gfunc$ and $\Gamma$. 
Since we are assuming that the chemical potential crosses only one band and the effective Hamiltonian describes that one band, $\Gamma, \Gfunc$ and $F$ in Eqs.~\eqref{eq:response_func},\eqref{eq:chi_gfunc_expression} are scalars (not matrices).

We can perform the summation over the Matsubara frequencies $\epsilon_n$ in Eq.~\eqref{eq:chi_gfunc_expression} with the standard technique and obtain 
\begin{widetext}
\begin{align}
	&\chi^{\mu\alpha\beta}(i\omegaoutput) = \int\frac{\dd{z}}{2\pi i} \int[\dd{k}] (-f(z))\Gamma^{\mu}(z+i\omegaoutput, z)\Gfunc(z + i\omegaoutput)F^{\alpha\beta}\Gfunc(z) \nonumber \\
	&= \int\frac{\dd{\ene}}{2\pi i} \int[\dd{k}] (-f(\ene))\{P^{\mu\alpha\beta}(\ene+i\eta, \ene+i\omegaoutput) - P^{\mu\alpha\beta}(\ene-i\eta, \ene+i\omegaoutput) + P^{\mu\alpha\beta}(\ene-i\omegaoutput, \ene+i\eta) - P^{\mu\alpha\beta}(\ene-i\omegaoutput, \ene-i\eta)\}
\end{align}
where $f(\ene) = (e^{\beta\ene}+1)^{-1}$ is the Fermi distribution function with the inverse temperature $\beta$, and $\eta$ is an infinitesimally small positive number. $P^{\mu\alpha\beta}(z_1, z_2)$ is defined as
\begin{align}
    P^{\mu\alpha\beta}(z_1, z_2) &= \Gamma^{\mu}(z_2, z_1)\Gfunc(z_2)F^{\alpha\beta}\Gfunc(z_1).
\end{align}
By analytically continuing the expression as $i\omegaoutput\to \omegaoutput+i\delta$, we obtain
\begin{align}
	&\chi^{\mu\alpha\beta}(\omegaoutput+i\delta) \nonumber\\
	&= \int\frac{\dd{\ene}}{2\pi i} \int[\dd{k}] \nonumber \\ 
	& \times (-f(\ene))\{P^{\mu\alpha\beta}(\ene+i\eta, \ene+\omegaoutput+i\delta) - P^{\mu\alpha\beta}(\ene-i\eta, \ene+\omegaoutput+i\delta) 
	+ P^{\mu\alpha\beta}(\ene-\omegaoutput-i\delta, \ene+i\eta) - P^{\mu\alpha\beta}(\ene-\omegaoutput-i\delta, \ene-i\eta)\} \label{eq:chi_w_vc1} \\
	&= \int\frac{\dd{\ene}}{2\pi i} \int[\dd{k}] \qty{(f(\ene)-f(\ene+\omegaoutput))P^{\mu\alpha\beta}(\ene-i\eta, \ene+\omegaoutput+i\delta) - f(\ene)\qty[P^{\mu\alpha\beta}(\ene+i\eta, \ene+\omegaoutput+i\delta) - P^{\mu\alpha\beta}(\ene-\omegaoutput-i\delta, \ene-i\eta)]}.
	 \label{eq:chi_w_vc2}
\end{align}
In particular, if we ignore the vertex correction, then $\Gamma^\mu$ becomes $h^\mu$ and the response function $\Phi^{\mu\alpha\beta}$ becomes
\begin{align}
    &\Phi^{\mu\alpha\beta}(\omegaoutput) \nonumber\\
    &= \int\frac{\dd{\ene}}{2\pi i} \int[\dd{k}] h^\mu F^{\alpha\beta}\qty[(f(\ene)-f(\ene+\omegaoutput))\GR(\ene+\omegaoutput)\GA(\ene) - f(\ene)(\GR(\ene+\omegaoutput)\GR(\ene)-\GA(\ene)\GA(\ene-\omegaoutput))] \nonumber \\
    & \quad + \int[\dd{k}] \expval{\partial_\mu F^{\alpha\beta}}_0, \label{eq:response_func_wo_vc} 
\end{align}

It should be noted that, without $\mathcal{T}$-breaking, $\Phi^{\mu\alpha\beta}(\omegaoutput)$ satisfies $\Phi^{\mu\alpha\beta}(\omegaoutput)=-\Phi^{\mu\beta\alpha}(\omegaoutput)$, and hence nonzero photocurrent responses require some form of $\mathcal{T}$-breaking such as a magnetic field, a magnetic order of the material, or application of the circularly polarized light.  

\subsection{Slow limit ($\omegaoutput\to0$ limit with finite scattering)} \label{subsec:finite_scattering_limit}
The case we are interested in is the dc response limit of $\omegaoutput\to0$. 
The first term in Eq.~\eqref{eq:chi_w_vc2} will vanish in this limit as long as $P^{\mu\alpha\beta}(\ene-i\eta, \ene+\omegaoutput+i\eta)$ does not diverge. For example, if we approximate the self energy as 
\begin{align}
	\selfene(i\ene) \simeq i\gamma\sgn(\ene) \label{eq:introduce_scattering_rate}
\end{align}
with a constant scattering rate $\gamma$ and ignore the vertex correction (see Eq.~\eqref{eq:response_func_wo_vc}),  $P^{\mu\alpha\beta}(\ene-i\eta, \ene+\omegaoutput+i\eta)$ in $\omegaoutput\to0$ limit becomes a Lorenztian as a function of $\ene$, where the peak is located at $\ene = \xi_{\vec{k}}$ with the width $\gamma$ and its peak value is $\propto 1/\gamma^2$,
while the factor $f(\ene)-f(\ene+\omegaoutput)$ effectively behaves as a window function with the width $k_B T$ and the height $\propto \omegaoutput$. Therefore, the first term in Eq.~\eqref{eq:response_func_wo_vc} behaves as $\propto(\omegaoutput/\gamma)(\min\{\kB T/\gamma, 1\})$.
Consequently, as long as $\omegaoutput\ll\gamma$, the first term vanishes in the $\omegaoutput\to 0$ limit. 

Then, Eq.~\eqref{eq:chi_w_vc2} reduces to the following form:
\begin{align}
	&\lim_{\omegaoutput\to0}\chi^{\mu\alpha\beta}(\omegaoutput+i\delta) \nonumber\\
	&= \int\frac{\dd{\ene}}{2\pi i} \int[\dd{k}] \qty(- f(\ene))\qty[P^{\mu\alpha\beta}(\ene+i\eta, \ene+i\delta) - P^{\mu\alpha\beta}(\ene-i\delta, \ene-i\eta)] \nonumber\\
	&= \int\frac{\dd{\ene}}{2\pi i} \int[\dd{k}] \qty(- f(\ene))\qty[\Gamma^{\mu}(\ene+i\eta, \ene+i\delta)\GR(\ene)F^{\alpha\beta}\GR(\ene) - \Gamma^{\mu}(\ene-i\delta, \ene-i\eta)\GA(\ene)F^{\alpha\beta}\GA(\ene)] \label{eq:equilibrium_term}.
\end{align}
\end{widetext}

It is useful to use the Ward-Takahashi identity to evaluate this quantity. The Ward-Takahashi identity states the vertex function $\Gamma^\mu(\ene\pm i\eta, \ene\pm i\delta)$ is related to the Green function as 
\begin{align}
	&\Gamma^\mu(\ene + i\eta, \ene + i\delta) = e\partial_{\mu}\qty(\xi_{\vec{k}}+ \selfeneR(\ene)), \\
	&\Gamma^{\mu}(\ene-i\delta, \ene-i\eta) = e\partial_{\mu}\qty(\xi_{\vec{k}} + \selfeneA(\ene)),
\end{align} 
where $\Sigma^{R/A}$ is the retarded/advanced self-energy. Therefore, we can prove the following relations between the vertex function and the Green functions:
\begin{align}
	&\GR(\ene)\Gamma^{\mu}(\ene+i\eta, \ene+i\delta)\GR(\ene) = e\partial_{\mu}(\GR), \\
	&\GA(\ene)\Gamma^{\mu}(\ene-i\delta, \ene-i\eta)\GA(\ene) = e\partial_{\mu}(\GA).
\end{align}
Note that $G^{R/A}$ is a number as we focus on a single band.
Inserting these relations into Eq.~\eqref{eq:equilibrium_term} and integrating by parts, we obtain
\begin{align}
	&\lim_{\omegaoutput\to0}\chi^{\mu\alpha\beta}(\omegaoutput+i\delta) \nonumber\\
	&= \int\frac{\dd{\ene}}{2\pi i} \int[\dd{k}] e f(\ene)\partial_{\mu}{F^{\alpha\beta}} \qty[\GR(\ene)-\GA(\ene)] 
\end{align} 
We note that under the approximation $\Im\GR(\ene) \simeq -\pi\delta(\ene-\tilde{\xi_{k}})$, we can rewrite 
\begin{align}
    &\lim_{\omegaoutput\to0}\chi^{\mu\alpha\beta}(\omegaoutput+i\delta) \nonumber\\
    &\simeq \int\frac{\dd{\ene}}{2\pi i} \int[\dd{k}] (-2\pi i \delta(\ene-\tilde{\xi}_{\vec{k}})) e f(\ene)\partial_{\mu}F^{\alpha\beta} \nonumber\\
    &= \int[\dd{k}] e (-f(\tilde{\xi}_{\vec{k}}))\partial_{\mu}F^{\alpha\beta}.
\end{align}
with the renormalized energy dispersion $\tilde{\xi}_{\vec{k}}$,
which implies its connection to the ``diamagnetic contribution'' $\propto \partial_{\mu}F^{\alpha\beta}$.

We can evaluate the diamagnetic term in Eq.~\eqref{eq:response_func} as well. The result is 
\begin{align}
	&\int[\dd{k}] \expval{\partial_\mu F^{\alpha\beta}}_0 \nonumber \\
	&= \int\frac{\dd{\ene}}{2\pi i} \int[\dd{k}] (-f(\ene))\partial_\mu F^{\alpha\beta} \qty[\GR(\ene)-\GA(\ene)] 
\end{align}
This exactly cancels the contribution from $\chi^{\mu\alpha\beta}$ and we find 
\begin{align}
	&\lim_{\omegaoutput\to 0}\Phi^{\mu\alpha\beta}(\omegaoutput+i\delta) = 0
\end{align}
This result means that the photocurrent vanishes if we properly take into account the effect of the scattering, which is consistent with Ref.~\cite{Belinicher1986, Ivchenko1988}.

\subsection{Fast limit and correspondence with the perturbation theory in the clean limit} \label{sec:cleanlimit_correspondence}
The results in the previous section does not coincide with the result from the perturbation theory in the clean limit. This is because the perturbation theory employed in, e.g., \cite{Parker2019}, assumes a different limit. To see this, we employ the approximation in Eq.~\eqref{eq:introduce_scattering_rate} and take the limit where $\omegaoutput$ and $\gamma$ approach zero under the condition $\omegaoutput\gg\gamma$, which we call the ``fast limit''. In other words, we first take $\gamma\to 0$ limit and then $\omegaoutput\to 0$ limit. 
In this case, the discussion at the beginning in Sec.~\ref{subsec:finite_scattering_limit} breaks down and the first term in Eq.~\eqref{eq:chi_w_vc2} does not vanish even if $\omegaoutput\to 0$ limit. If we neglect the vertex correction, the response function in this limit is 
\begin{widetext}
\begin{align}
	\Phi^{\mu\alpha\beta}(\omegaoutput+i\delta) &= \int\frac{\dd{\ene}}{2\pi i} \int[\dd{k}] (f(\ene)-f(\ene+\omegaoutput))eh^{\mu}F^{\alpha\beta}\GR(\ene+\omegaoutput)\GA(\ene) \nonumber\\
	&= \int\frac{\dd{\ene}}{2\pi i} \int[\dd{k}] (f(\ene)-f(\ene+\omegaoutput))eh^{\mu}F^{\alpha\beta}\frac{1}{-\omegaoutput-2i\gamma}\qty(\frac{1}{\ene+\omegaoutput-\xi_{\vec{k}} + i\gamma}-\frac{1}{\ene-\xi_{\vec{k}}-i\gamma}) \nonumber\\
	&\to_{\gamma\to 0} \int\frac{\dd{\ene}}{2\pi i} \int[\dd{k}] \frac{f(\ene+\omegaoutput)-f(\ene)}{\omegaoutput}eh^{\mu}F^{\alpha\beta}\qty(\frac{1}{\ene+\omegaoutput-\xi_{\vec{k}} + i\eta}-\frac{1}{\ene-\xi_{\vec{k}}-i\eta}) \nonumber \\
	&\to_{\omegaoutput\to0} \int\frac{\dd{\ene}}{2\pi i} \int[\dd{k}] f'(\ene) eh^{\mu}F^{\alpha\beta}\qty(\frac{1}{\ene-\xi_{\vec{k}} + i\eta}-\frac{1}{\ene-\xi_{\vec{k}}-i\eta}) \nonumber \\
	&= \int[\dd{k}] (-f'(\xi_{\vec{k}})) eh^{\mu}F^{\alpha\beta}= \int[\dd{k}] e\qty(-\partial_{\mu}f(\xi_{\vec{k}})) F^{\alpha\beta} \nonumber \\
	&= \int[\dd{k}] e^3 \qty(-\partial_{\mu}f(\xi_{\vec{k}}))  \qty[h^{\alpha\beta}_{aa} + \sum_{b\in\highband} \qty(\frac{h^{\alpha}_{ab}h^{\beta}_{ba}}{\omega_{ab}-\omega} + \frac{h^{\beta}_{ab}h^{\alpha}_{ba}}{\omega_{ab}+\omega})]
	\label{eq:results_fastlimit}
    \end{align}
\end{widetext}
where we replaced $\gamma$ with an infinitesimal positive quantity $\eta$ when taking $\gamma\to0$ limit.

Noting the relation Eq.~\eqref{eq:b_EE_relation}, the result of the fast limit, Eq.~\eqref{eq:results_fastlimit}, coincides with the result from the perturbation theory in the clean limit (Eq.~\eqref{eq:perturbation_results}). 

\subsection{Crossover from the fast limit to the slow lmiit}
We can also discuss the crossover region from the fast limit ($\omegaoutput\gg\gamma$) to the slow limit ($\omegaoutput\ll\gamma$), i.e., $\omegaoutput\sim\gamma$, with Eq.~\eqref{eq:chi_w_vc2}. As an example, we consider a generalized Rice-Mele model. The Hamiltonian of the Rice-Mele model is defined as 
\begin{align}
    \hat{H} &= \sum_{n} \qty(t_{AB} \cre{c}_{n,B} \ani{c}_{n,A} + t_{BA}\cre{c}_{n+1, A}\ani{c}_{n,B} + h.c.) \nonumber \\
	&+ \sum_{n} \qty(t_{AA} \cre{c}_{n+1, A}\ani{c}_{n,A} + t_{BB} \cre{c}_{n+1, B}\ani{c}_{n,B} + h.c.) \nonumber \\
	&+ \sum_{n} \frac{\Delta}{2}\qty(\cre{c}_{n,A}\ani{c}_{n,A} - \cre{c}_{n,B}\ani{c}_{n,B}). \label{eq:Rice-Mele}
\end{align}
Here $i$ is the index for unit cells, $A$ and $B$ are labels for the two sites in a unit cell, and their positions in the unit cell are denoted by $r_A, r_B$. Then the Fourier transformation of $\ani{c}_{n,i}$ for $i=A,B$ is defined as 
\begin{align}
    \ani{c}_{k,i} &= \sum_{n} e^{-ik(na+r_i)} \ani{c}_{n,i},
\end{align}
and the Hamiltonian $\hat{H}$ can be rewritten as  
\begin{align}
    &\hat{H} = \sum_{k} \begin{bmatrix}
        \cre{c}_{k,A} & \cre{c}_{k,B}
    \end{bmatrix} H_0(k)
    \begin{bmatrix}
        \ani{c}_{k,A} \\
        \ani{c}_{k,B}
    \end{bmatrix}, \\
    &H_0(k) = \begin{bmatrix}
        \frac{\Delta}{2} + t_{AA} \cos{ka} & t(k) \\
        t(k)^{*} & -\frac{\Delta}{2} + t_{BB} \cos{ka}
    \end{bmatrix}, \\
    &t(k) = t_{AB}^{*}e^{-ik(r_A-r_B)} + t_{BA} e^{-ik(a + r_A-r_B)},
\end{align}
with the size of unit cell $a$.

\begin{figure*}[htbp]
    \centering
    \includegraphics[width=2.0\columnwidth]{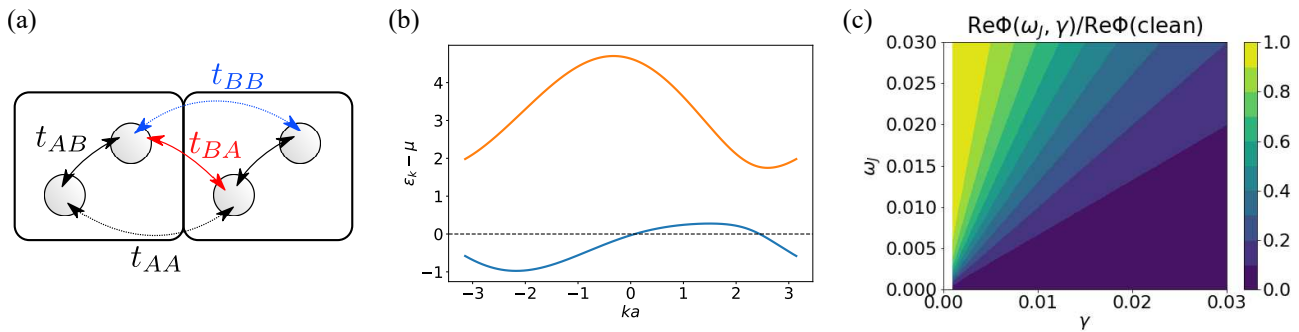}
    \caption{Photocurrent in Rice-Mele model with time reversal symmetry broken. (a) The schematic illustration of the Rice-Mele model (see also Eq.~\eqref{eq:Rice-Mele}). (b) The band structure of the Rice-Mele model. The used parameters are $t_{AB}=1+0.5i$, $t_{BA}=1-0.5i$, $t_{AA}=0.5$, $t_{BB}=0.3$, $r_A=0, r_B=0.5$, $\Delta=2.0$, $\mu=-1.5$, $T=0.05$. The size of unit cell $a$ is set to be $1$. (c) The photocurrent response of the Rice-Mele model for the incident photon frequency $\omega=0.5$, which is smaller than the direct optical gap $E_g\sim1$. Eq.~\eqref{eq:response_func_wo_vc} is calculated numerically for $0\le\omegaoutput\le0.03, 0.001\le\gamma\le0.03$. The vertex correction is ignored in the calculation. The calculated value is normalized by the value in the fast limit, i.e., Eq.~\eqref{eq:results_fastlimit}. }
    \label{fig:crossover_RM}
\end{figure*}

One can introduce the inversion symmetry breaking into the model by setting, for example, $\Delta\neq 0$ and $\abs{t_{AB}}\neq \abs{t_{BA}}$. We further break the time reversal symmetry $\cal{T}$ by introducing a complex hopping. The second-nearest-neighbor hopping is necessary to break $\cal{T}$-symmetry; otherwise the effect of the complex hopping can be eliminated by a gauge transformation. We show a schematic picture of the Rice-Mele model in Fig.~\ref{fig:crossover_RM}(a). We calculated the band structure (Fig.~\ref{fig:crossover_RM}(b)) and the real part of the nonlinear response function $\Phi^{\mu\alpha\beta}(\omegaoutput)$ (Fig.~\ref{fig:crossover_RM}) for the parameters given in the caption of Fig.~\ref{fig:crossover_RM}. $\Phi^{\mu\alpha\beta}(\omegaoutput)$ is calculated as a function of $\omegaoutput$ and $\gamma$ without vertex correction by calculating Eq.~\eqref{eq:response_func_wo_vc} numerically.

Figure~\ref{fig:crossover_RM}(c) shows the nonlinear response function as a function of $\gamma$ and $\omegaoutput$ and a crossover between the fast limit and the slow limit.
One can see that for the slow limit $\omegaoutput\ll\gamma$, $\mathrm{Re}\Phi^{\mu\alpha\beta}$ is almost zero and hence the photocurrent response vanishes as we showed in Sec.~\ref{subsec:finite_scattering_limit}. On the other hand, in the region $\omegaoutput\gg\gamma$, i.e., the fast limit, the photocurrent response is nonzero and approaching to the value in the fast limit, which is consistent with the results in Sec~\ref{sec:cleanlimit_correspondence}.

\section{Discussion and conclusion} \label{sec:discussion}
First we note that $\Phi^{\mu\alpha\beta}$ is finite only when the time reversal symmetry $\mathcal{T}$ is broken either by magnetic field/order of the system or circularly polarized light. Since the photocurrent response is allowed only when the spatial inversion symmetry $\mathcal{P}$ is broken, both breaking $\mathcal{T}$ and $\mathcal{P}$ is important for the photocurrent discussed in the present paper.

Our discussion clarifies the effects of scattering on the photocurrent. If the scattering rate $\gamma$ is sufficiently large compared to the frequency of the output photocurrent $\omegaoutput$, the system supports photocurrent proportional to $\omegaoutput$ at the steady state, and the photocurrent vanishes in the limit $\omegaoutput/\gamma\to 0$. This can be understood as a realization of an ``equilibrium'' state with renomalized energy spectrum~\cite{Belinicher1986, Ivchenko1988}. Indeed, since the system is described by the effective Hamiltonian \eqref{eq:Heff}, the perturbation is effectively static for the slow limit $\omegaoutput\ll\gamma$, and an ``equilibrium'' state with the effective Hamiltonian is realized, hence no current occurs. This discussion generalizes a semiclassical result by Belinicher {\it et al.}~\cite{Belinicher1986} to a quantum theoretical treatment. 

The effective Hamiltonian derived in this work is closely related to an effect called dynamical Stark effect or optical Stark effect as well~\cite{Autlert1955, Bakos1977, Sie2015, Sie2017, Pershoguba2022a, Pershoguba2022arXiv}, where a shift of energy levels is induced by an external ac field. Indeed, the dc component of Eq.~\eqref{eq:Heff} coincides with the expression for optical Stark shift~\cite{Pershoguba2022arXiv}. In Ref.~\cite{Pershoguba2022arXiv}, the dc current induced by the applied ac electric fields with frequency smaller than the band gap is also discussed, and it is concluded that only transient current before the system relaxes into an ``equilibrium'' state is possible. 

We also note that, as a transient current, there would be another contribution, described by $\pdv{P_{\mathrm{tr}}^{\mu}}{t}$ in Eq.~\eqref{eq:Jeff_maintext}. This contribution may be interpreted as a polarization current (see Appendix \ref{ap:effective_Hamiltonian} for details), and may be related to the optical rectification effect, where the electric polarization is induced by ac electric fields~\cite{Bass1962, Bass1965}. 

On the other hand, the perturbation theory in the clean limit gives finite photocurrent, as we reviewed in Sec.~\ref{sec:review}. This limit is recovered in the fast limit ($\gamma\ll\omegaoutput$), as we have seen in Sec.~\ref{sec:cleanlimit_correspondence}. Physically, this limit represents the relaxation rate of the system $\gamma$ is sufficiently small compared to the rate of extracting electrons from the system to outside the sample (e.g., to the electrodes attached to the sample). Although usually $\omegaoutput\ll\gamma$ holds for dc photocurrent, this limit may be realized in some situations. For example, if the system is thin in the current direction compared to the scattering length, the electron carrying the current quickly flows out of the system before scattering happens, where the effective scattering rate becomes almost zero. In this case, the $\omegaoutput\gg\gamma$ limit is effectively realized and nonzero photocurrent within transparent region may be possible. Similar situation is experimentally realized in BaTiO$_3$ thin film~\cite{Spanier2016, Zenkevich2014}, although $\mathcal{T}$ is preserved in BaTiO$_3$. In similar setup with $\mathcal{T}$ breaking materials or circularly-polarized light, one may observe finite photocurrent within transparent region. 

\begin{figure}
    \centering
    \includegraphics[width=\columnwidth]{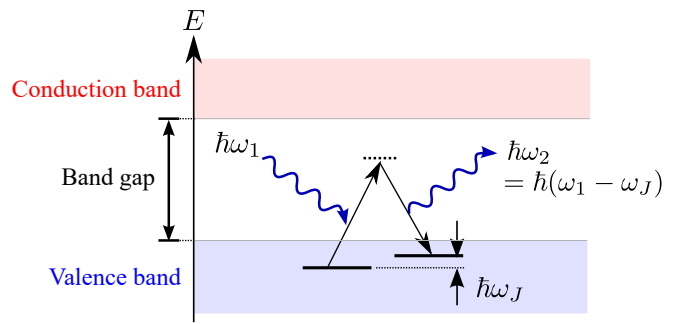}
    \caption{Interpretation of the photocurrent within the transparent region as a Raman-like process. The photocurrent for the fast limit (Eq.~\eqref{eq:results_fastlimit}) can be understood as a Raman-like process, where the photon with frequency $\omega_1$ is scattered by the sample but with a different frequency $\omega_2=\omega_1-\omegaoutput$, leaving an excitation of energy $\hbar\omegaoutput$. Here, $\omega_1,\omega_2\simeq\omega$ and $\omegaoutput\ll\omega$. The excitation with energy $\hbar\omegaoutput$ results in a finite current. However, this current will be canceled if the relaxation rate $\gamma$ is sufficiently fast compared with $\omegaoutput$, i.e., in the slow limit.}
    \label{fig:schematics_Raman}
\end{figure}

The finite current in the fast limit can be interpreted as a Raman process as well. Namely, the photon with frequency $\omega_1$ is scattered by the sample, and the scattered photon has a frequency $\omega_2=\omega_1-\omegaoutput$, leaving an excitation of energy $\hbar\omegaoutput$ with a finite current in the sample (Fig.~\ref{fig:schematics_Raman}). Here $\omega_1,\omega_2\simeq\omega$ and $\omegaoutput\ll\omega$. Since the energy of the incident photon is smaller than the band gap, an electron is only virtually excited to the conduction band from the valence band during this process. However, in the presence of finite scattering or relaxation (i.e., the slow limit), the induced current will be canceled. 

$\omegaoutput\gg\gamma$ limit may be related to the recent study by Shi {\it et al.}~\cite{Shi2022arXiv}. They discussed systems coupled with particle reservoirs. The coupling to the reservoirs results in the relaxation rate $\Gamma$, and it is shown that finite photocurrent is induced when $\Gamma$ is finite. Here, we note that the relaxation rate $\Gamma$ in Ref.~\cite{Shi2022arXiv} has essentially a different meaning from what we denote as $\gamma$ in the present paper. $\Gamma$ in Ref.~\cite{Shi2022arXiv} represents in a sense the particle exchange rate with the particle reservoir, while $\gamma$ in the present work is the scattering rate within the system. Therefore, they can have different effects on the photocurrent behavior, and the results in Ref.~\cite{Shi2022arXiv} do not contradict to our results. 

Our discussion does not take into account these scattering effects during a virtual excitation since we start from the effective Hamiltonian \eqref{eq:Heff}. Therefore, those scattering effects may enable finite photocurrent within the transparent region. Indeed, Golub {\it et al.} recently proposed that finite photocurrent can be induced even within the transparent region if one considers the impurity scattering during a virtual excitation~\cite{Golub2022arXiv}. Their proposal might look similar to our interpretation of the current in the fast limit as a Raman process, but it is essentially different because they are considering the scattering effect during a virtual excitation of electrons and they also discussed the finite momentum of the light.

In conclusion, we have derived the effective Hamiltonian under illumination of light with frequency smaller than the band gap, and clarify the effects of relaxation on photovoltaic effect. We show that if the frequency of the output current $\omegaoutput$ is much smaller than the scattering rate $\gamma$, the photocurrent will be proportional to $\omegaoutput/\gamma$ and vanishes in the limit of $\omegaoutput/\gamma\to 0$.
In contrast,if the condition $\omegaoutput\gg\gamma$ is met in some situations, nonzero photocurrent can appear in the transparent region. 

{\it Note added}: After submitting the first version of this manuscript to arXiv, we received many comments and revised the manuscript to the current version substantially.

\begin{acknowledgements}
The authors acknowledge the useful discussions with Inti Sodemann and Li-kun Shi which 
inspired and motivated this work. The authors are grateful for the useful comments by L.E. Golub M.M. Glazov, and V. Yakovenko. 
We also thank Y. Tokura, and N. Ogawa for useful discussion. 
This work was supported by JSPS KAKENHI Grant
[No.~JP18H03676 (N.N.), No.~JP21J00453 (H.W.) and No.~JP22J22111 (Y.O.)],
JST PRESTO [Grant No.~JPMJPR19L9 (T.M.)],
and
JST CREST [Grant No.~JPMJCR19T3 (T.M.) and JPMJCR1874 (N.N.)].
\end{acknowledgements}

\appendix

\section{Construction of the effective Hamiltonian} \label{ap:effective_Hamiltonian}
In this appendix, we construct the effective Hamiltonian for noninteracting electronic systems under ac electric fields, the frequency of which is smaller than the band gap. To this end, we utilize the Schrieffer-Wolff transformation for the time-dependent Hamiltonian.

\subsection{Model}
Let us consider the system under the ac electric field $\vec{E}(t)$. The Bloch Hamiltonian in the band basis is given by 
\begin{align}
	H(t) &= \Emat - eA^\alpha(t) h^\alpha + \frac{e^2}{2}A^{\alpha}(t) A^{\beta}(t) h^{\alpha\beta} + \order{A^3}, 
\end{align}
where $\vec{A}(t)$ is the vector potential and $e(<0)$ is the charge of the electron. $\alpha, \beta$ are spatial indices and the Einstein convention is always used in this paper. Here and hereafter, the planck constant $\hbar$ is set to be 1.

The vector potential $\vec{A}(t)$ is related to the electric field $\vec{E}(t)$ as $\vec{E}(t) = -\dv{\vec{A}}{t}$. We further assume that the vector potential is given by 
\begin{align}
	\vec{A}(t) &= \vec{a}(t)e^{-i\omega t} + (\vec{a}(t))^{*} e^{i\omega t},
\end{align}
where $\vec{a}(t)$ varies slowly compared to the frequency $\omega$.

\subsection{Time-dependent Schrieffer-Wolff transformation}
Here we would like to find an effective Hamiltonian for low energy bands. We denote the set of the low energy bands as $\lowband$, and the other bands as $\highband$. 

To construct an effective Hamiltonian, let us consider a time dependent unitary transformation $U(t)$. $U(t)$ transforms the Hamiltonian $H(t)$ into $H'(t)$ as 
\begin{align}
	H'(t) &= U(t)H(t)U^{\dagger}(t) - i U(t)\partial_tU^{\dagger}(t).
\end{align}
We determine $U(t)$ so that $H'(t)$ is block-diagonal up to $\order{A}$, i.e., 
\begin{align}
	H'(t)_{ab} = \order{A^2} \quad \text{for $a/b\in\lowband$ and $b/a\in\highband$} \label{eq:condition_linear}
\end{align}

We further assume that $U(t)$ is of the form $U(t) = \exp(S(t))$ with $S(t)=\order{A}$. Then the new Hamiltonian $H'(t)$ becomes 
\begin{align}
	H'(t) &= \Emat - eA^{\alpha}(t)h^{\alpha} + \comm{S(t)}{\Emat} + i\partial_t S(t) \nonumber \\
	& + \frac{e^2}{2}A^{\alpha}(t)A^{\beta}(t) h^{\alpha\beta} + \comm{S(t)}{-eA^{\alpha}(t)h^{\alpha}} + \frac{1}{2}\acomm{S(t)^2}{\Emat} \nonumber \\
	& - S(t)\Emat S(t) + \frac{i}{2}\comm{S(t)}{\partial_t S(t)} + \order{A^3}.
\end{align}
Thus the condition \eqref{eq:condition_linear} becomes
\begin{align}
	(-eA^\alpha(t)h^{\alpha} + \comm{S(t)}{\Emat} + i\partial_t S(t))_{ab}=0
\end{align}
for $a/b\in\lowband, b/a\in\highband$. This can be rewritten as 
\begin{align}
	(\ene_{ba} + i\partial_t)S_{ab} = e A^{\alpha}(t) h^{\alpha}_{ab},
\end{align}
where $\ene_{ab}=\ene_a-\ene_b$. 
By using the Fourier transformation, this equation can be easily solved and we obtain 
\begin{align}
	S_{ab}(t) &= \int\frac{\dd{\Omega}}{2\pi} e^{-i\Omega t} \frac{eA^\alpha(\Omega)h^\alpha_{ab}}{\Omega-\ene_{ab}} \nonumber\\
	&\simeq eh^{\alpha}_{ab}\qty(\frac{a^\alpha(t)e^{-i\omega t}}{\omega-\ene_{ab}} + \frac{(a^{\alpha}(t))^{*}e^{i\omega t}}{-\omega-\ene_{ab}}) \label{eq:Sab_expression}
\end{align}
for $a/b\in\lowband, b/a\in\highband$. Here, $A^{\alpha}(\Omega)$ is the Fourier transformation of $A^{\alpha}(t)$. In the derivation of Eq.~\eqref{eq:Sab_expression}, we ignore the time dependence of $\vec{a}(t)$. The other components of $S$ is 0.

Then $H'_{aa'}$ for $a, a'\in\lowband$ becomes
\begin{align}
	H'_{aa'}(t) &\simeq \delta_{aa'}\ene_a - eA^{\alpha}(t)h^{\alpha}_{aa'} + b^{\alpha\beta}(t) F^{\alpha\beta}_{aa'} \nonumber\\
	&\quad + (\text{terms with $e^{-2i\omega t}, e^{2i\omega t}$}) + \order{A^3}\\
	F^{\alpha\beta}_{aa'} &= e^2 h_{aa'}^{\alpha\beta} + \frac{e^2}{2} \sum_{b\in\highband}\left[\qty(\frac{1}{\ene_{ab}-\omega} + \frac{1}{\ene_{a'b}-\omega})h^{\alpha}_{ab}h^{\beta}_{ba'} \right. \nonumber \\
	& \quad \left. + \qty(\frac{1}{\ene_{ab}+\omega} + \frac{1}{\ene_{a'b}+\omega})h^{\beta}_{ab}h^{\alpha}_{ba'}\right], \\
	b^{\alpha\beta}(t) &= a^{\alpha}(t)(a^\beta(t))^{*},
\end{align}
and for $a/b \in \lowband$, $b/a\in\highband$, 
\begin{align}
	H'_{ab}(t) &= \frac{e^2}{2}A^\alpha(t)A^{\beta}(t) h^{\alpha\beta}_{ab} \nonumber \\
	& - eA^\alpha(t)\sum_c \left[\qty(\frac{a^{\beta}(t)e^{-i\omega t}}{\omega-\ene_{ac}} + \frac{(a^{\beta}(t))^*e^{i\omega t}}{-\omega-\ene_{ac}})eh^\beta_{ac}h^\alpha_{cb} \right. \nonumber\\
	& \left. - eh^\alpha_{ac}\qty(\frac{a^{\beta}(t)e^{-i\omega t}}{\omega-\ene_{cb}} + \frac{(a^{\beta}(t))^*e^{i\omega t}}{-\omega-\ene_{cb}})h^\beta_{cb}\right]
\end{align}

To eliminate the $\order{A^2}$ terms in off-diagonal matrix elements in $H'$, we further apply another unitary transformation $V(t) = \exp(T(t))$ with $T(t)=\order{A^2}$ to $H'(t)$ to obtain $H''(t)$. $H''(t)$ is given by 
\begin{align}
	H''(t) &= H'(t) + \comm{T(t)}{H'(t)} + i\partial_t T(t) + \order{A^3},
\end{align}
and the condition to determine $T(t)$ is 
\begin{align}
	0 &= H''(t)_{ab} \nonumber \\
	&= H'_{ab} + T_{ab}(H'_{bb}-H'_{aa}) + i\partial_t T_{ab}. \label{eq:condition_T}
\end{align}
Note that $T=\order{A^2}$ and $T_{aa'}=0$ for $a,a'\in\lowband$. Using the solution for Eq.~\eqref{eq:condition_T}, the new Hamiltonian $H''_{aa'}(t)$ for $a, a'\in\lowband$ becomes 
\begin{align}
	H''_{aa'}(t) &= H'_{aa'}(t) + \sum_{b\in\highband}\qty(T_{ab}H'_{ba'}-H'_{ab}T_{ba'}) \nonumber\\
	&\quad + \sum_{a''\in\lowband}\qty(T_{aa''}H'_{a''a'}-H'_{aa''}T_{a''a'}) + i\partial_t T_{aa'}\nonumber \\
	&\quad + \order{A^3} \nonumber \\
	&= H'_{aa'}(t) + \order{A^3}.
\end{align}
Therefore, $H''_{aa'}(t)$ coincides with $H'_{aa'}(t)$.

\subsection{Transformed current operator}
Since we are interested in the current response, we will need the current operator after the Schrieffer-Wolff transformation. The current operator before the transformation is given by 
\begin{align}
	J^\mu(t) &= \partial_\mu H(t) - i\comm{\bc^{\mu}}{H(t)},
\end{align}
where $\bc^{\mu}$ is the interband Berry connection. 

The new current operator $J''^\mu(t)$ is given by
\begin{align}
	J''^\mu(t) &= \tilde{U}(t) J^{\mu}(t) \tilde{U}^{\dagger}(t) \nonumber \\
	&= \tilde{U}(t) (\partial_\mu H(t) - i\comm{\bc^{\mu}}{H(t)}) \tilde{U}^{\dagger}(t) \nonumber\\
	&= \partial_\mu H''(t) - i\comm{\bc''^\mu(t)+\zeta''^\mu(t)}{H''(t)} 
	  \nonumber \\
	& \quad + \partial_\mu\zeta''_t(t) - i\comm{\bc''^\mu(t)+\zeta''^\mu(t)}{\zeta''_t(t)}, \label{eq:effective_current_operator}
\end{align}
where $\tilde{U}(t)=V(t)U(t), \bc''^\mu(t) = \tilde{U}(t) \bc^{\mu} \tilde{U}^{\dagger}(t)$ and 
\begin{align}
	\zeta''^\mu(t) &= i\tilde{U}(t) \partial_\mu\tilde{U}^{\dagger}(t), \\
	\zeta''_t(t) &= i\tilde{U}(t) \partial_t\tilde{U}^{\dagger}(t).
\end{align}
The third and fourth terms in Eq.~\eqref{eq:effective_current_operator} can be rewritten as a total derivative with respect to time:
\begin{align}
    & \partial_\mu\zeta''_t(t) - i\comm{\bc''^\mu(t)+\zeta''^\mu(t)}{\zeta''_t(t)} = \pdv{P_{\mathrm{tr}}^\mu}{t}, \\
    & P_{\mathrm{tr}}^\mu = \bc''^\mu(t)+\zeta''^\mu(t),
\end{align}
and hence this is a transient current. $P_{\mathrm{tr}}$ may be interpreted as a polarization current~\cite{Resta1994, Vanderbilt1993}. Indeed, $\bc''^\mu$ is the Berry connection after the unitary transformation $\tilde{U}(t)$, and $\zeta''^\mu$ is also a Berry connection-like quantity defined with $\tilde{U}(t)$, instead of the unitary transformation diagonalizing the Bloch Hamiltonian without external fields. 

In particular, $J''^\mu_{aa'}(t)$ for $a,a'\in\lowband$ is given by 
\begin{align}
	J''^\mu_{aa'}(t) &= \partial_\mu H''_{aa'}(t) \nonumber \\
	& \quad - i\sum_{a''\in\lowband} \left[\qty(\bc''^\mu_{aa''}(t)+\zeta''^\mu_{aa''}(t))H''_{a''a'}(t) \right. \nonumber\\
	& \quad \left. -H''_{aa''}(t)\qty(\bc''^\mu_{a''a'}(t)+\zeta''^\mu_{a''a'}(t))\right] + \pdv{P^\mu_{\mathrm{tr},aa'}}{t} + \order{A^3}.
\end{align}
Here we used $H''_{ab}(t)=\order{A^3}$ for $a\in\lowband$ and $b\in\highband$.
Furthermore, since $H''_{aa'}(t)=\order{A}$ if $a\neq a'$ for $a, a'\in\lowband$ and $\zeta''^\mu_{aa'}(t) = \order{A^2}$ for $a,a'\in\lowband$, $J''^{\mu}_{aa'}(t)$ is simplified to 
\begin{align}
	J''^\mu_{aa'}(t) &= \partial_\mu H''_{aa'}(t) \nonumber\\
	& \quad - i\sum_{a''\in\lowband} \qty(\bc''^\mu_{aa''}(t)H''_{a''a'}(t)-H''_{aa''}(t)\bc''^\mu_{a''a'}(t)) \nonumber \\
	& \quad - i\zeta''^\mu_{aa'}(t)\ene_{a'a} + \pdv{P_{\mathrm{tr},aa'}^\mu}{t} + \order{A^3} \label{eq:current_operator}
\end{align}
If we denote the projection operator onto $\lowband$ as $\proj$, Eq.~\eqref{eq:current_operator} can be rewritten as 
\begin{align}
	&\proj J''^\mu(t)\proj \nonumber \\
	\quad &= \partial_\mu \proj H''(t)\proj - i\comm{\proj\bc''^\mu \proj}{\proj H''(t)\proj} \nonumber\\
	\quad & \quad - i\comm{\proj\zeta''^\mu \proj}{\proj \Emat \proj}+ \proj\pdv{P_{\mathrm{tr}}^\mu}{t}\proj+\order{A^3} \label{eq:current_operator_with_P}
\end{align}

In particular, if $\lowband$ includes only one band, the second and the third terms in Eq.~\eqref{eq:current_operator_with_P} vanish and we obtain
\begin{align}
	\proj J''^\mu(t)\proj &= \partial_\mu \proj H''(t) \proj + \proj\pdv{P_{\mathrm{tr}}^\mu}{t}\proj,
\end{align}
which corresponds to Eq.~\eqref{eq:Heff} in the main text. As in the main text, if we neglect the transient current $\pdv{P^\mu}{t}$ and the term linear in $A^{\alpha}(t)$, the effective current operator is given by
\begin{align}
    \proj J''^\mu(t)\proj &= \partial_{\mu}\ene_a + b^{\alpha\beta}\partial_\mu F^{\alpha\beta}_{aa},
\end{align}
where $a$ is the index for the band in $\lowband$.

In summary, we have obtained an effective Hamiltonian and the corresponding current operator as 
\begin{align}
	&\proj H''(t) \proj = \proj\Emat \proj - eA^{\alpha}(t)\proj h^{\alpha}\proj  + b^{\alpha\beta}(t) F^{\alpha\beta} \\
	&F^{\alpha\beta}_{aa'} = e^2 h_{aa'}^{\alpha\beta} + \frac{e^2}{2} \sum_{b\in\highband}\left[\qty(\frac{1}{\ene_{ab}-\omega} + \frac{1}{\ene_{a'b}-\omega})h^{\alpha}_{ab}h^{\beta}_{ba'} \right. \nonumber \\
	& \quad\quad \left. + \qty(\frac{1}{\ene_{ab}+\omega} + \frac{1}{\ene_{a'b}+\omega})h^{\beta}_{ab}h^{\alpha}_{ba'}\right], \\
	& \proj J''^\mu(t)\proj  = \partial_\mu \proj H''(t)\proj  - i\comm{\proj \bc''^\mu \proj }{\proj H''(t)\proj } \nonumber\\
	& \quad - i\comm{\proj \zeta''^\mu \proj }{\proj  \Emat \proj }+ \proj\pdv{P_{\mathrm{tr}}^\mu}{t}\proj+ \order{A^3}
\end{align}
In the case that $\lowband$ consists of only one band with index $a$, these results are simplified to the following:
\begin{align}
	\Heff(t) &= \ene_a - eA^{\alpha}(t)h^{\alpha}_{aa} + F^{\alpha\beta}b^{\alpha\beta}(t), \\
	F^{\alpha\beta} &= e^2 h^{\alpha\beta}_{aa} + e^2\sum_{b\neq a} \qty[\frac{h^{\alpha}_{ab}h^{\beta}_{ba}}{\ene_{ab}-\omega} + \frac{h^{\beta}_{ab}h^{\alpha}_{ba}}{\ene_{ab}+\omega}],\\
	\Jeff^{\mu}(t) &= \partial_\mu \Heff(t) + \pdv{P_{\mathrm{tr}}^{\mu}}{t}, \\
	A^\alpha(t) &= a^{\alpha}(t) e^{-i\omega t} + (a^{\alpha}(t))^{*} e^{i\omega t}, \\
	b^{\alpha\beta}(t) &= a^{\alpha}(t)(a^{\beta}(t))^{*},
\end{align}

\bibliography{library, library_arXiv}

\end{document}